\documentclass[]{spie}  

\usepackage{amsmath,amsfonts,amssymb}
\usepackage{graphicx}
\usepackage[colorlinks=true, allcolors=blue]{hyperref}
\usepackage{tabu}
\DeclareMathOperator*{\argmax}{arg\,max}

\usepackage{physics}

\title{Approximate Graph Spectral Decomposition with the Variational Quantum Eigensolver}

\author[a]{Josh Payne}
\author[a]{Mario Srouji}
\affil[a]{Department of Computer Science\\Stanford University}

\authorinfo{Further author information: (Send correspondence to J.P.)\\J.P.: E-mail: jfp@cs.stanford.edu\\  M.S.: E-mail: msrouji@stanford.edu}

\begin{document} 
\maketitle

\begin{abstract}
Spectral graph theory is a branch of mathematics that studies the relationships between the eigenvectors and eigenvalues of Laplacian and adjacency matrices and their associated graphs. The Variational Quantum Eigensolver (VQE) algorithm was proposed as a hybrid quantum/classical algorithm that is used to quickly determine the ground state of a Hamiltonian, and more generally, the lowest eigenvalue of a matrix $M\in \mathbb{R}^{n\times n}$. There are many interesting problems associated with the spectral decompositions of associated matrices, such as partitioning, embedding, and the determination of other properties. In this paper, we will expand upon the VQE algorithm to analyze the spectra of directed and undirected graphs. We evaluate runtime and accuracy comparisons (empirically and theoretically) between different choices of ansatz parameters, graph sizes, graph densities, and matrix types, and demonstrate the effectiveness of our approach on Rigetti's QCS platform on graphs of up to 64 vertices, finding eigenvalues of adjacency and Laplacian matrices. We finally make direct comparisons to classical performance with the Quantum Virtual Machine (QVM) in the appendix, observing a superpolynomial runtime improvement of our algorithm when run using a quantum computer.\footnote{Code: \url{https://github.com/Josh-Payne/Quantum-Graph-Spectra/}}
\end{abstract}


\section{Introduction}
\subsection{Preliminaries} Quantum computing is an emerging paradigm in computation which leverages the quantum mechanical phenomena of superposition and entanglement to create states that scale exponentially with number of qubits, or quantum bits. Quantum algorithms have been proposed which have considerable speedups in a wide variety of algebraic and number theoretic problems such as factoring of large numbers \cite{Shor} and matrix multiplication \cite{buhrman}. In 2013 Peruzzo et al. proposed a variational quantum eigenvalue solver, which was targeted towards finding the ground state of a Hamiltonian, specifically of a quantum chemical system. \cite{Peruzzo} In this report, we will extend this work to analyze the spectra of the matrices associated with \textit{graphs}, which are mathematical structures that denote relationships (via \textit{edges}) between objects (via \textit{vertices}). 
\subsection{The Adjacency and the Laplacian}
For a graph $G$ with vertex set $V$, the adjacency matrix is a square $|V| \times |V|$ matrix $A(G)$ such that its element $A(G)_{ij} = 1$ when there is an edge from vertex i to vertex j, and $0$ when there is no edge. The Laplacian matrix is a square $|V| \times |V|$ matrix $L(G)$ such that its element $L(G)_{ij} = -1$ when there is an edge from vertex i to vertex j, $0$ when there is no edge, and $L(G)_{ii} = deg(v_i)$, where $v_i$ is the $i^{th}$ vertex in $V$. If the graph is directed, $L(G)_{ii}$ may correspond to the indegree or outdegree of $v_i$; we will explore both. \cite{biggs}
\\ We will be working with directed and undirected graphs without self-loops. $\lambda$ is an \textit{eigenvalue} if for some nonzero vector $x$ and matrix $A$, $Ax = \lambda x$. We order the eigenvalues of any $n\times n$ adjacency matrix as $\lambda_{min} = \lambda_n\leq \ldots \leq \lambda_1 = \lambda_{max}$, and the eigenvalues of an $n\times n$ Laplacian matrix as $\mu_{min} = \mu_n\leq \ldots \leq \mu_1 = \mu_{max}$. We quickly note that $\mu_{min} = 0$, so we will be more interested in finding $\mu_{max}$.

\subsection{Variational Quantum Eigensolver (VQE)}
The Variational Quantum Eigensolver (VQE) algorithm combines the ability of quantum computers to efficiently compute expectation values with a classical optimization routine in order to approximate ground state energies of quantum systems. VQE allows us to find the smallest eigenvalue  $\lambda_n$ (and corresponding eigenvector) of a matrix $A$. It is based on the variational principle, which states that for any Hamiltonian H,
$$\frac{\bra{\psi}H\ket{\psi}}{\bra{\psi}\ket{\psi}} \ge \lambda_n,$$
and $$\frac{\bra{\psi}(-H)\ket{\psi}}{\bra{\psi}\ket{\psi}} \ge -\lambda_1.$$ There are two subroutines involved with VQE. The quantum subroutine has two steps: first, we prepare an $ansatz$, or a quantum state $\ket{\psi(\theta)} $ parameterized by $\theta \in [0,2\pi]^n$. Then, we measure the expectation value $$\mathbb{E}\left[\frac{\bra{\psi(\theta)}H\ket{\psi(\theta)}}{\bra{\psi(\theta)}\ket{\psi(\theta)}}\right].$$
The classical subroutine is as follows. We use a classical non-linear optimizer such as the Nelder-Mead method \cite{nm} to minimize the expectation value by varying the ansatz parameters $\theta$. Then, we iterate this step until convergence. It has been shown that VQE demonstrates polynomial scaling of each iteration with respect to system size, in contrast to exponential scaling of the current best-known classical algorithm for the same task. Further, the update step of the Nelder-Mead method scales linearly (or, at worst case, polynomially) in the number of parameters included in the minimization of the expectation of the above term \cite{fletcher}. Let $M$ be the number of terms comprising the Hamiltonian $H$, $p$ be the desired precision value, $h_{max} = \displaystyle\argmax_h(h\langle\sigma\rangle)$ where where $h$ is a constant and $\sigma$ is the $k$-fold tensor product of Pauli operators acting on the system, and $n$ be such that $H\in\mathbb{R}^{2^n\times 2^n}$. The authors of \cite{Peruzzo} estimate the total cost per iteration to be $O(n^r|h_{max}|^2M/p^2)$, for some small constant $r$ which is determined by the encoding of the quantum state and the classical minimization method (in our case, Nelder-Mead). In contrast, the computation of the expectation value $\langle\sigma\rangle$ = $\bra{\psi}\sigma\ket{\psi}$ using classical methods requires $O(2^n)$ floating point operations, giving that the scaling of this procedure for a classical computer is roughly $O(M2^{n(r+1)})$, which demonstrates a superpolynomial quantum speedup over the classical alternative. Note that we are not claiming a superpolynomial speedup over a method for finding spectra, as the authors of \cite{fast_LA} and others have shown that polynomial-time algorithms for this exist. We are rather claiming that the method described is more efficient on a quantum processor, and has practical applications, as argued by the authors of \cite{Peruzzo}.

\subsection{Spectral Graph Theory Applications}
We can explore multiple problems in spectral graph theory by representing the adjacency and Laplacian matrices of graphs as Pauli operators, and then leveraging the VQE algorithm to find the minimum (or maximum) eigenvalues and respective eigenvectors of these matrices. Note that given the largest eigenvalue $\lambda_1$ and corresponding eigenvector $v_1$ of a symmetric matrix $M$ corresponding to the adjacency or Laplacian of an undirected graph, the largest eigenvalue of $M'=M-\lambda_1 \frac{v_1 v_1^T}{\| v_1 \|^2}$ is the second-largest eigenvalue of $M$, since $v_1$ now has eigenvalue 0. Using this fact we can decompose the spectra of these matrices. This, in turn, will reveal interesting properties about the graphs. Here are a few applications, though the focus of our paper is on the process of gathering the eigenvalues:
\begin{enumerate}
    \item If we take the sum of the squares of the distances between neighbors, then the eigenvector corresponding to the smallest non-zero eigenvalue will be minimizing tfhis sum of squared distances. Likewise, the maximum eigenvalue $\max_{||v||=1}\displaystyle\sum_{(i,j)\in E, \ i<j} (v(i) - v(j))^2$ will try to maximize the discrepancy between neighbors’ values of v. \footnote{https://web.stanford.edu/class/cs168/l/l11.pdf}
    \item The eigenvectors corresponding to small eigenvalues are, in some sense, trying to find good partitions of a graph. These low eigenvectors are trying to find ways of assigning different numbers to vertices, such that neighbors have similar values. Additionally, since they are all orthogonal, each eigenvector is trying to find a “different” or “new” such partition. 
    \item Many problems can be modeled as the problem of k-coloring a graph (assigning one of k colors to each vertex in a graph, where no two neighboring vertices have the same color). This problem of finding a k-color, or even deciding whether a k-coloring of a graph exists, is NP-hard in general. One natural heuristic is to embed the graph onto the eigenvectors corresponding to the highest eigenvalues. As one would expect, in these embeddings, points that are close together in the embedding tend to not be neighbors in the original graph. \cite{sgt}
    \item $\lambda_1 = 2d \iff G$ is bipartite.
\end{enumerate}

\section{Proposed Method}
\subsection{Pauli Representation}
We are given an adjacency or Laplacian matrix of a graph, and want to represent it as an operation of Pauli matrices. To do this, we recursively break up a $2^n\times2^n$ matrix into $2\times2$ submatrices, and can consequently represent any matrix in $\mathbb{R}^{2^n\times2^n}$ as a Pauli sum/product. This allows us to represent any graph on $2^n$ vertices, where some of the vertices can be used for padding. The procedure is as follows.
\\ We may represent any matrix $M_2 \in \mathbb{R}^{2\times2}$ as a linear combination of the below constructor matrices $C^{(i)}$:
$$C^{(1)} := \begin{pmatrix} 1 & 0 \\ 0 & 0 \end{pmatrix} = \frac{1}{2}(I_0 + Z_0)$$
$$C^{(2)} := \begin{pmatrix} 0 & 1 \\ 0 & 0 \end{pmatrix} = \frac{1}{2}(X_0 + Z_0X_0)$$
$$C^{(3)} := \begin{pmatrix} 0 & 0 \\ 1 & 0 \end{pmatrix} = \frac{1}{2}(X_0 - Z_0X_0)$$
$$C^{(4)} := \begin{pmatrix} 0 & 0 \\ 0 & 1 \end{pmatrix} = \frac{1}{2}(I_0 - Z_0)$$

We may then represent any $2^n\times2^n$ adjacency matrix $M_{2^n}$ = $\begin{pmatrix} A_{2^{n-1}} & B_{2^{n-1}} \\ C_{2^{n-1}} & D_{2^{n-1}} \end{pmatrix}$ as Pauli operators by representing its submatrices $A_{2^{n-1}}$, $B_{2^{n-1}}$, $C_{2^{n-1}}$, and $D_{2^{n-1}} \in \mathbb{F}_2^{2^{n-1}\times2^{n-1}}$ as Pauli operators. Note that $$M_{2^n} = C^{(1)}\otimes A_{2^{n-1}} + C^{(2)}\otimes B_{2^{n-1}} + C^{(3)}\otimes C_{2^{n-1}} + C^{(4)}\otimes D_{2^{n-1}},$$ where we can construct $A_{2^{n-1}}$, $B_{2^{n-1}}$, $C_{2^{n-1}}$, and $D_{2^{n-1}}$ recursively, where the base case is a linear combination of the $C^{(i)}$.

To represent a Laplacian matrix an operation of Pauli matrices, let $P(A)$ be the Pauli operation representing the adjacency matrix, and $P(L)$ be the Pauli operation representing the Laplacian matrix. Then $$P(L) = -P(A)+\displaystyle\sum_{i=1}^{2^n}deg(v_i)P(\mathbf{1}_i),$$ where $\mathbf{1}_i$ is $2^n\times2^n$ matrix which is $1$ at ($i$, $i$) and $0$ everywhere else. $deg(v_i)$ may be specified to be the indegree or outdegree of $v_i$ in the case of a directed graph.

For each matrix type, we may represent a graph on $|V|$ vertices by embedding it into a $2^{\lceil\log_2(|V|)\rceil}\times2^{\lceil\log_2(|V|)\rceil}$ matrix and padding the unused rows and columns with zeros, as this does not affect the spectra. Resulting $n$-fold tensored Pauli operator construction is then simplified using the Pauli algebra rules. 

\subsection{Eigenvalue Estimation}
\begin{figure}[t]
    \centering
\includegraphics[width=0.5\textwidth]{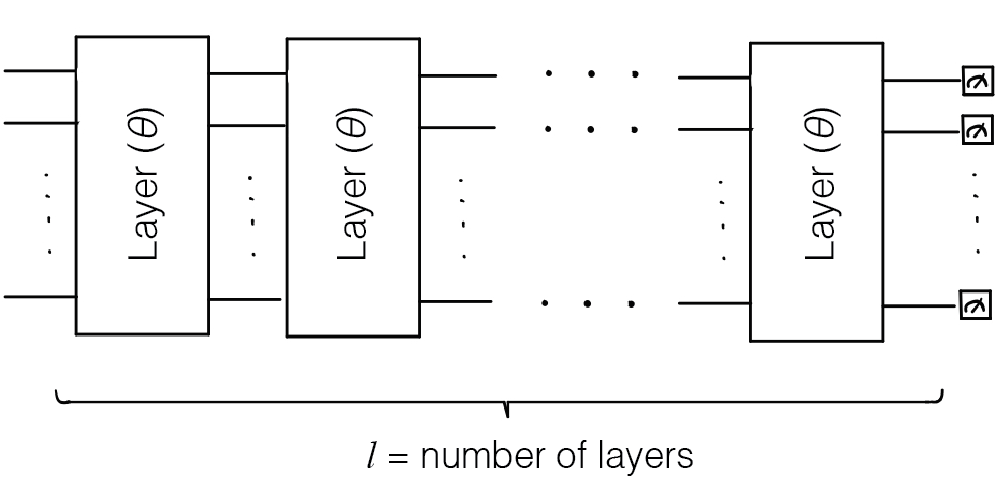}
\caption{Concatenated Layers in our Ansatz}
    \label{fig:wholeans}
\end{figure}

\begin{figure}[t]
    \centering
\includegraphics[width=0.5\textwidth]{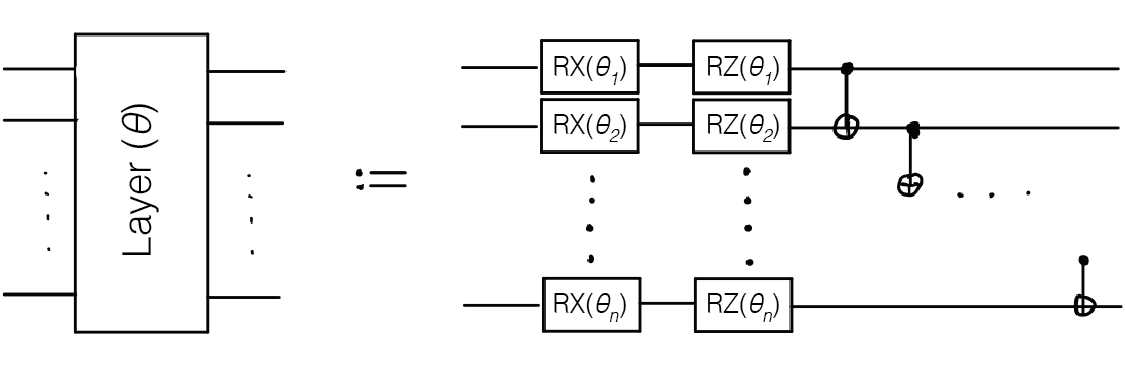}
\caption{One Layer of our Ansatz}
    \label{fig:layer}
\end{figure}

Once we've represented our matrix as an operation of Pauli matrices, we can utilize the Variational Quantum Eigensolver to determine the minimum eigenvalue of the system. For the purposes of this experiment, we opted to use the layered ansatz proposed by \cite{kandala}. Finding improved ansatzes that perform well specifically for matrices related to graphs is left as future work. The ansatz can be represented as follows. In \ref{fig:wholeans}, we concatenate parameterized layers of gates some number of times to produce an ansatz of a given depth $l$. 
Now, let $n$ be the number of qubits our ansatz is applied to. Apply \texttt{RX}$(\theta_i)$ followed by \texttt{RZ}$(\theta_i)$ to every qubit. Then apply \texttt{CNOT}$(q, q+1)$ for $q \in \{0,...,n-1\}$. This applies \texttt{CNOT} gates to entangle all of our qubits after doing some rotation that is parameterized by $\theta$. This is seen in \ref{fig:layer}.
. The number of layers in the ansatz is a hyperparameter: the more layers that are present, the more computationally intensive the procedure is, but also generally the more accurate the estimation is. The Variational Quantum Eigensolver takes as input the Pauli representation of our matrix, acting on $\lceil{\log_2(|V|)}\rceil$ qubits, as each tensor operation in our Hamiltonian construction step multiplies the number of rows and columns by 2.
With this chosen ansatz, we opted to use the Nelder-Mead method referenced in 1.C.

\section{Experimental Results and Analyses}
\begin{figure}[t]
    \centering
\includegraphics[width=0.5\textwidth]{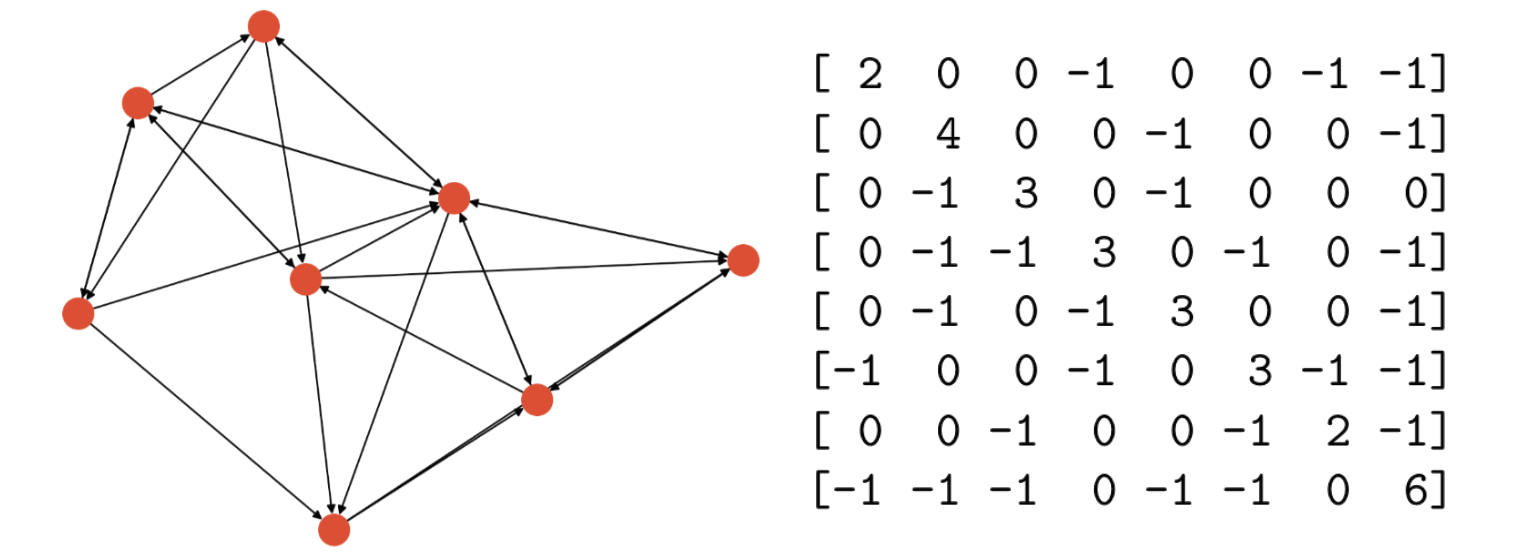} 
\caption{Directed graph on 8 vertices and its Laplacian indegree matrix}
    \label{fig:graphnmat}
\end{figure}

\begin{figure}[t]
    \centering
\includegraphics[width=0.5\textwidth]{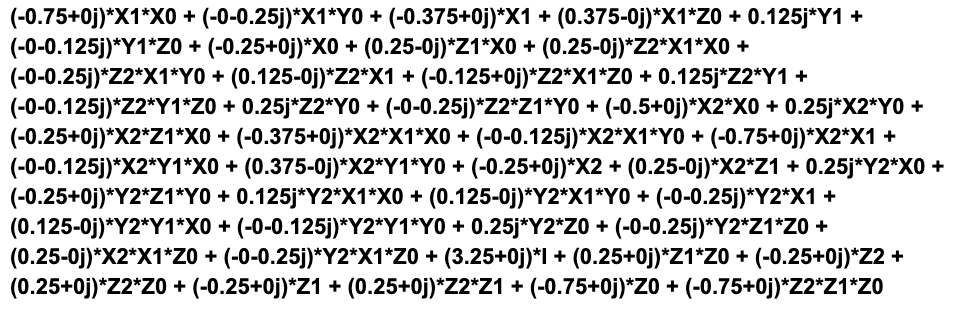}
\caption{Pauli Operations}
    \label{fig:pauli}
\end{figure}

To evaluate our approach, we chose to implement the algorithms in pyQuil, a library for generating Quil programs to be executed using the Rigetti Forest platform. Quil is a quantum instruction set architecture that first introduced a shared quantum/classical memory model. \cite{quil} We first tested our algorithms on the Quantum Virtual Machine (QVM) before running them on Rigetti's Quantum Cloud Service (QCS) platform, on lattices of varying topologies. For a comprehensive evaluation, we analyzed runtime and calculation error (by taking the absolute value of the difference between our output and the output of \texttt{numpy}'s \texttt{linalg.eig} function) comparisons between different choices of ansatz parameters, graph densities, and matrix types on graphs of between 4 and 64 vertices. In \ref{fig:graphnmat}, \ref{fig:graphnmat}, and \ref{fig:pauli}, we see an example of a directed graph on 8 vertices, its (indegree) Laplacian, and the corresponding Pauli operator term.
\subsection{Gate Complexity}
A corollary to what was shown by Stuart Hadfield in \cite{hadfield} is that if a function $f_n$ is \textit{efficiently representable} as a Hamiltonian $H_{f_n}$, then \texttt{size}$(H_{f_n})$ is \texttt{poly}$(n)$. Since $f_n$ is a $k$-fold tensor product with $k$ growing on the order of $\log(n)$, it is efficiently representable. Moreover, Hadfield showed that the number of gates needed to represent $H_{f_n}$ is $O(deg(H_f)\texttt{size}(H_f))$, where $deg(H_f)$ is the maximum locality of any term (number of qubits the term acts on), so we conjecture that the complexity of the gates in our system is \texttt{poly}$(n)$, as well.

To test this, we randomly generated adjacency matrices undirected graphs of density $0.5$ with numbers of vertices being 4, 8, 16, 32, 64, 128, and 256, and selected the resulting largest Pauli representations of each size to gather ``worst-average case" values on the number of gates with respect to the input size. We then plotted these values and fitted a curve to test our conjecture.
\begin{figure}[t]
    \centering
\includegraphics[width=0.5\textwidth]{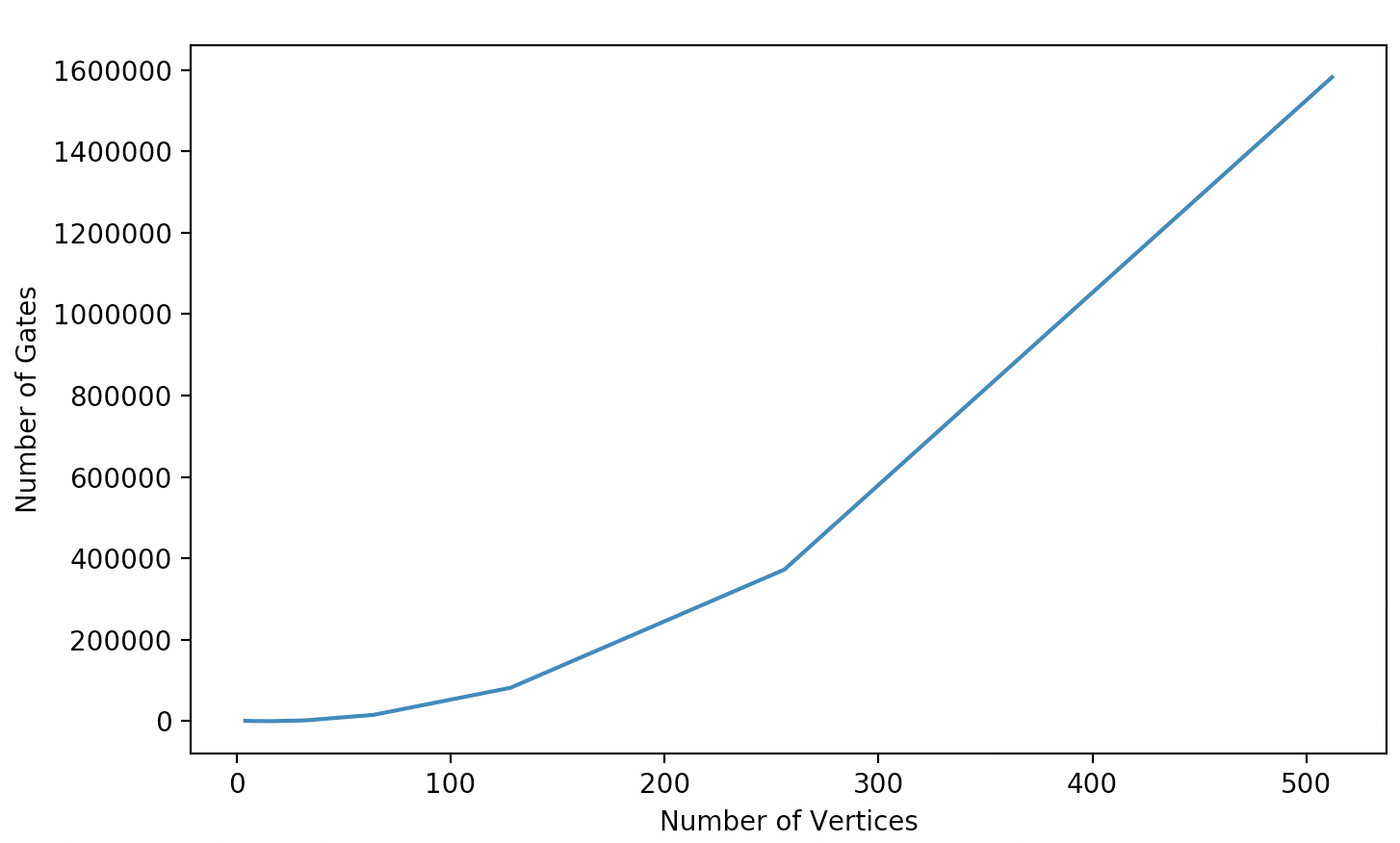}
\caption{Gate Complexity}
    \label{fig:gate}
\end{figure}

The results in \ref{fig:gate} seem to support our conjecture, as we were able to fit a quadratic curve to the data.

\subsection{Graph Densities}
Before we tested our eigenvalue estimation algorithms on QCS with respect to any other parameter, we wanted to determine the worst-case density of a graph. We tested the accuracy and runtime for our algorithm with 36 densities spaced uniformly between 0 and 1. The other parameters are as follows:
\\
\\
\begin{center}
\begin{tabu} to 0.5\textwidth { | X[c] | X[c] | }
 \hline
 Matrix Type & Undirected Adjacency \\
 \hline
 Number of Vertices & 8  \\
 \hline
 Number of Trials per Test & 3  \\
 \hline
 Number of Ansatz Layers & 3  \\
\hline
 Lattice & Aspen-4-3Q-A  \\
\hline
\end{tabu}
\end{center}
\begin{figure}[t]
    \centering
\includegraphics[width=0.5\textwidth]{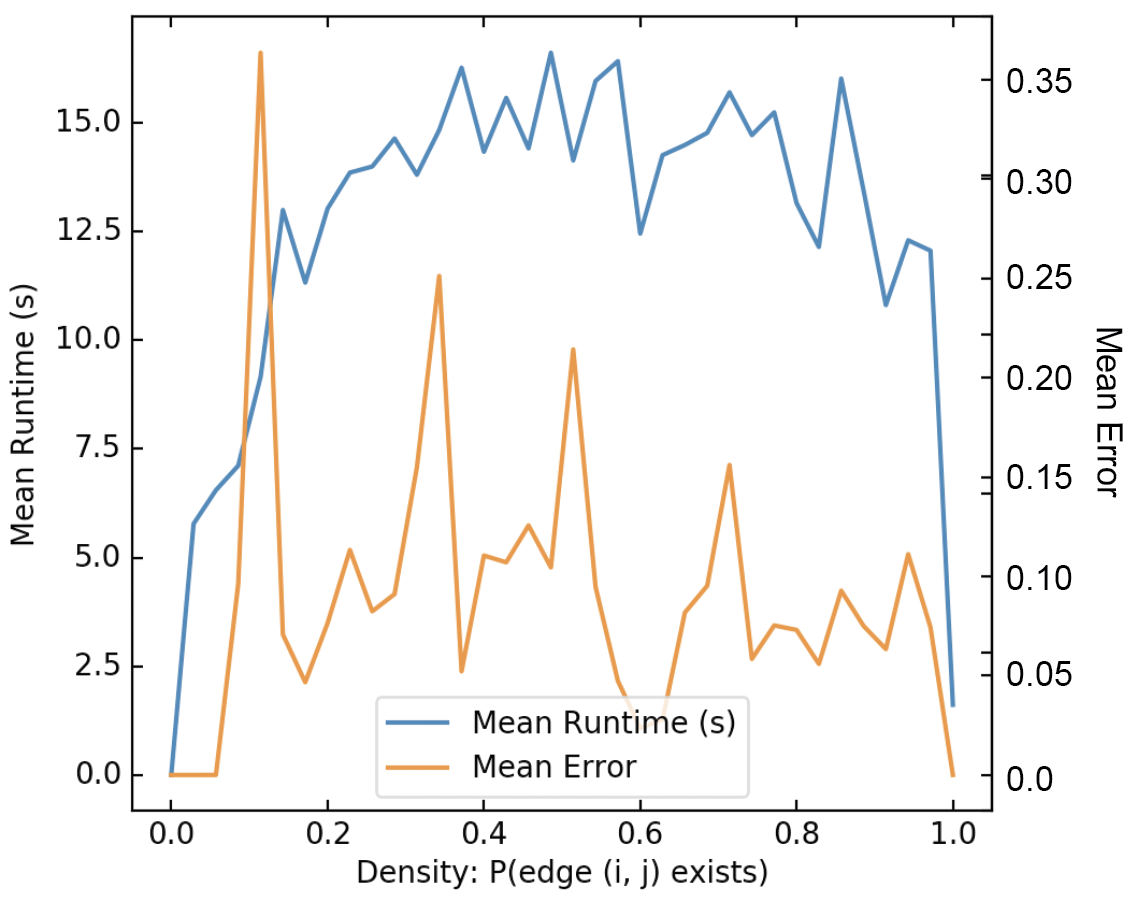}
\caption{Runtime and Error vs. Density}
    \label{fig:redens}
\end{figure}
As we expected, we see in \ref{fig:redens} that the algorithm performed very quickly and with low error on ``trivial" graphs, those that were fully connected or empty. The runtime peaks with ``half-connected" graphs, where the density parameter is $0.5$. As a result, we chose this value as the density parameter for most of our other tests. For good measure to reduce variance, we ran this test with more examples on densities of 0, 0.1, 0.3, 0.5, 0.7, 0.9, and 1:
\\
\\
\begin{center}
\begin{tabu} to 0.5\textwidth { | X[c] | X[c] | }
 \hline
 Matrix Type & Undirected Adjacency \\
 \hline
 Number of Vertices & 4  \\
 \hline
 Number of Trials per Test & 20  \\
 \hline
 Number of Ansatz Layers & 5  \\
\hline
 Lattice & Aspen-4-2Q-A  \\
\hline
\end{tabu}
\begin{figure}[t]
    \centering
 \includegraphics[width=0.5\textwidth]{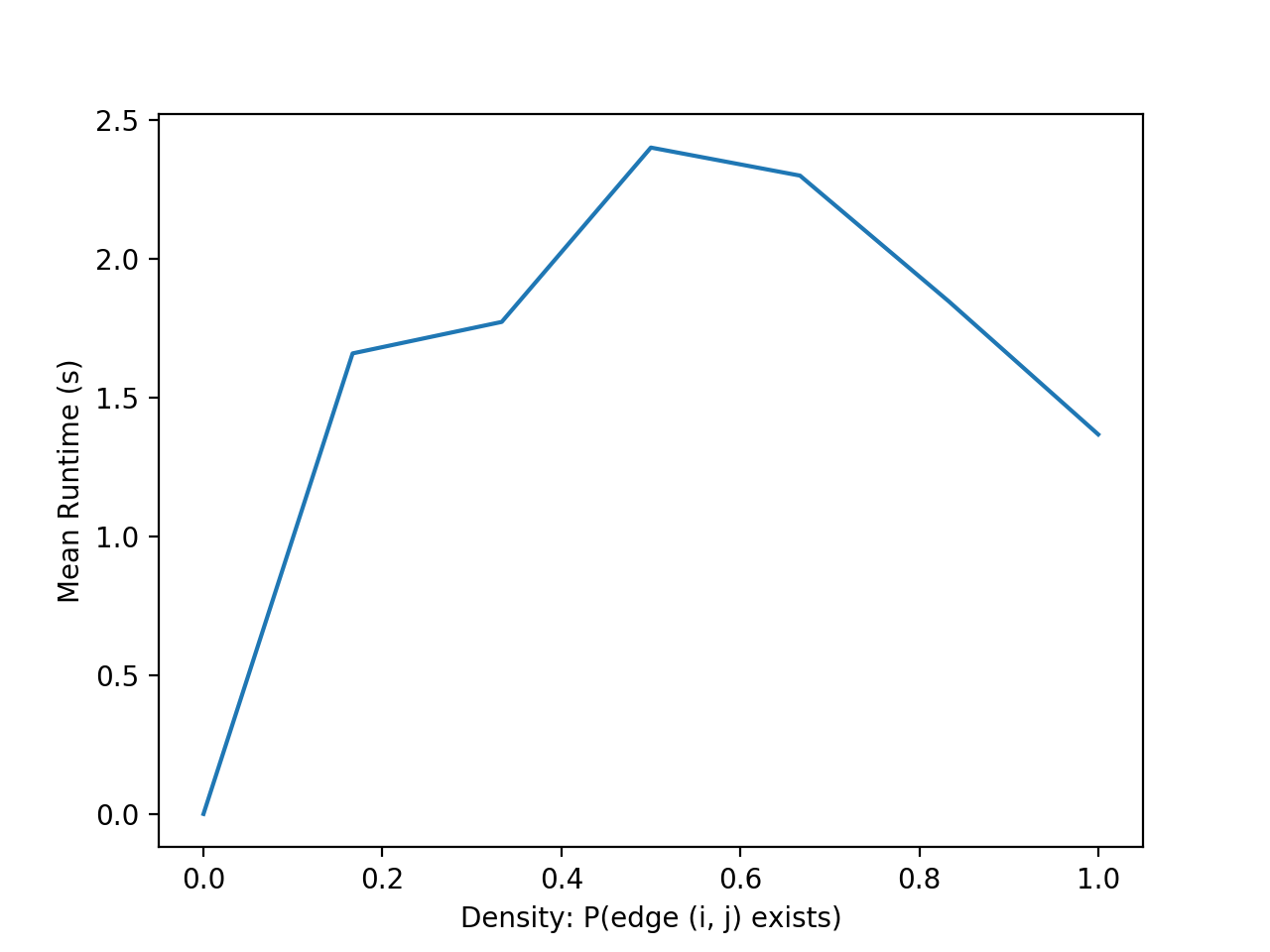}
 \caption{Runtime vs. Density}
    \label{fig:dens}
\end{figure}
\end{center}
\ref{fig:dens} supports our previous conclusion about the relationship between density and runtime. 

\subsection{Ansatz Layers}
We next wanted to determine the effect that the number of layers, $l$, of our ansatz had on the runtime and accuracy of our algorithm. We chose to test ansatzes of 1, 2, 3, 4, 5, 7, 10, 15, and 20 layers.
\\
\\
\begin{center}
\begin{tabu} to 0.5\textwidth { | X[c] | X[c] | }
 \hline
 Matrix Type & Undirected Adjacency \\
 \hline
 Number of Vertices & 4  \\
 \hline
 Number of Trials per Test & 20  \\
 \hline
 Density & 0.5  \\
\hline
 Lattice & Aspen-4-2Q-A  \\
\hline
\end{tabu}
\end{center}

\begin{figure}[t]
    \centering
 \includegraphics[width=0.5\textwidth]{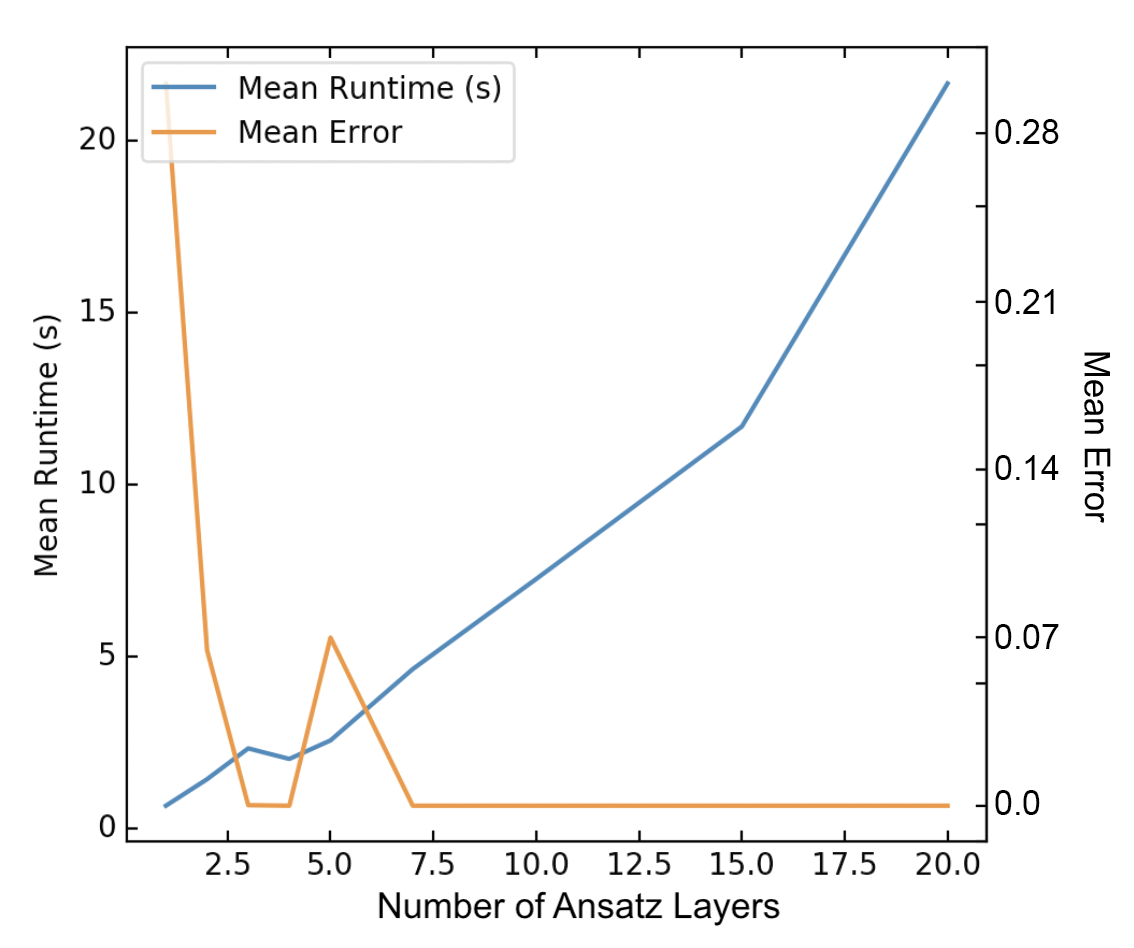} 
 \caption{Runtime vs. Ansatz Layers}
    \label{fig:ans}
\end{figure}
As we suspected, the error rate seen in \ref{fig:ans} decreased as the number of ansatz layers increased, but the computation time also increased. We noted that at around 3 layers, there was a point of diminishing returns in terms of accuracy for 4-vertex graph matrices, so we chose to use 3-layer ansatzes in future experiments.
\subsection{Graph and Matrix Types}
Next, we wanted to test the effectiveness of our algorithm with respect to different graph and matrix types. In particular, we wanted to fix other variables and determine the runtime difference between the adjacency matrix of an undirected graph, the adjacency matrix of a directed graph, the Laplacian matrix of an undirected graph, the Laplacian matrix of an undirected graph (with outdegree), and the Laplacian matrix of an undirected graph (with indegree). In these tests, we found the maximum eigenvalue of each respective matrix, since the minimum eigenvalue of the Laplacian is $0$.
\\
\\
\begin{center}
\begin{tabu} to 0.5\textwidth { | X[c] | X[c] | }
 \hline
 Number of Vertices & 8  \\
 \hline
 Number of Trials per Test & 5  \\
 \hline
 Density & 0.5  \\
  \hline
 Number of Ansatz Layers & 3  \\
\hline
 Lattice & Aspen-4-3Q-A  \\
\hline
\end{tabu}
\end{center}
\begin{figure}[t]
    \centering
 \includegraphics[width=0.5\textwidth]{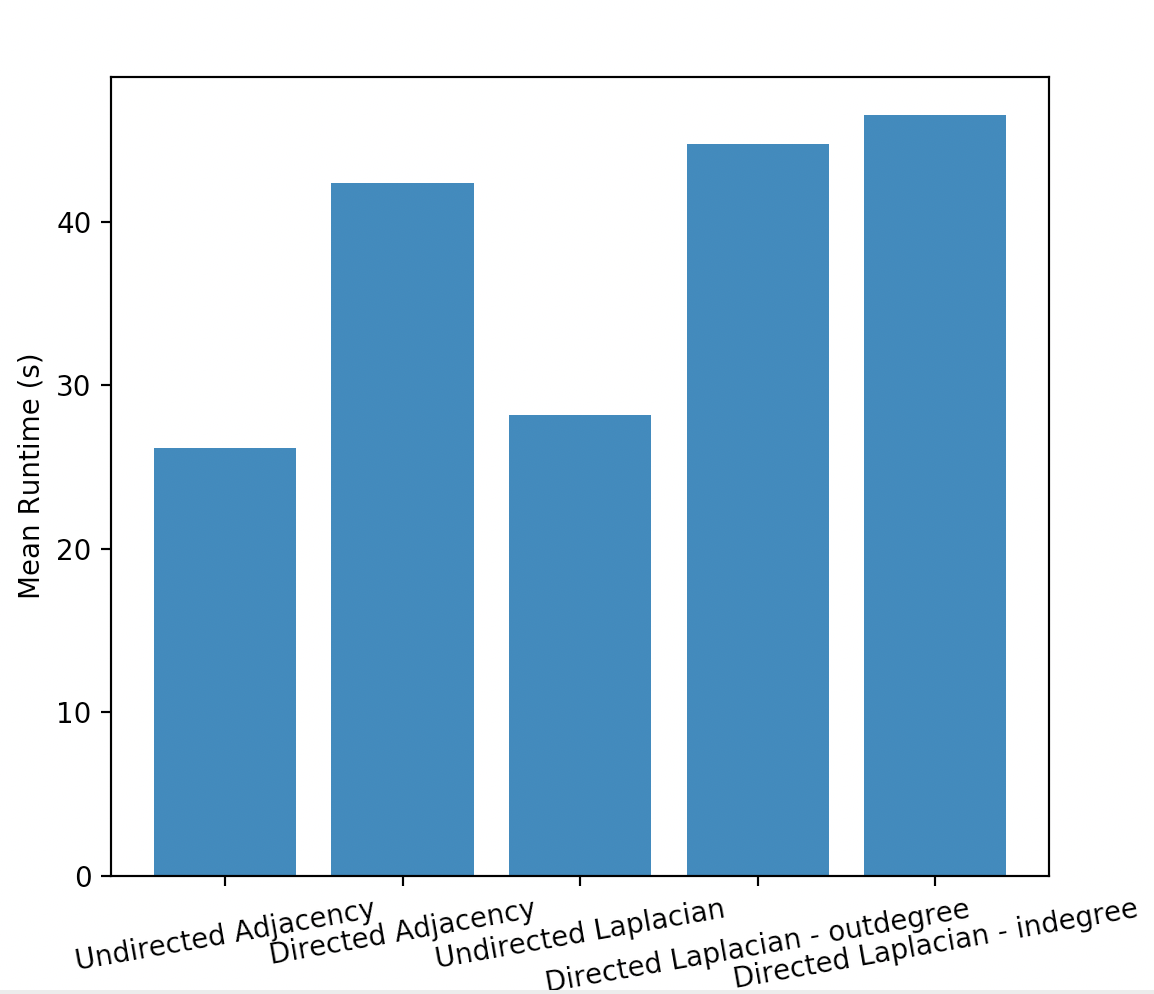}
 \caption{Matrix Type vs. Runtime}
    \label{fig:typer}
\end{figure}

\begin{figure}[t]
    \centering
 \includegraphics[width=0.5\textwidth]{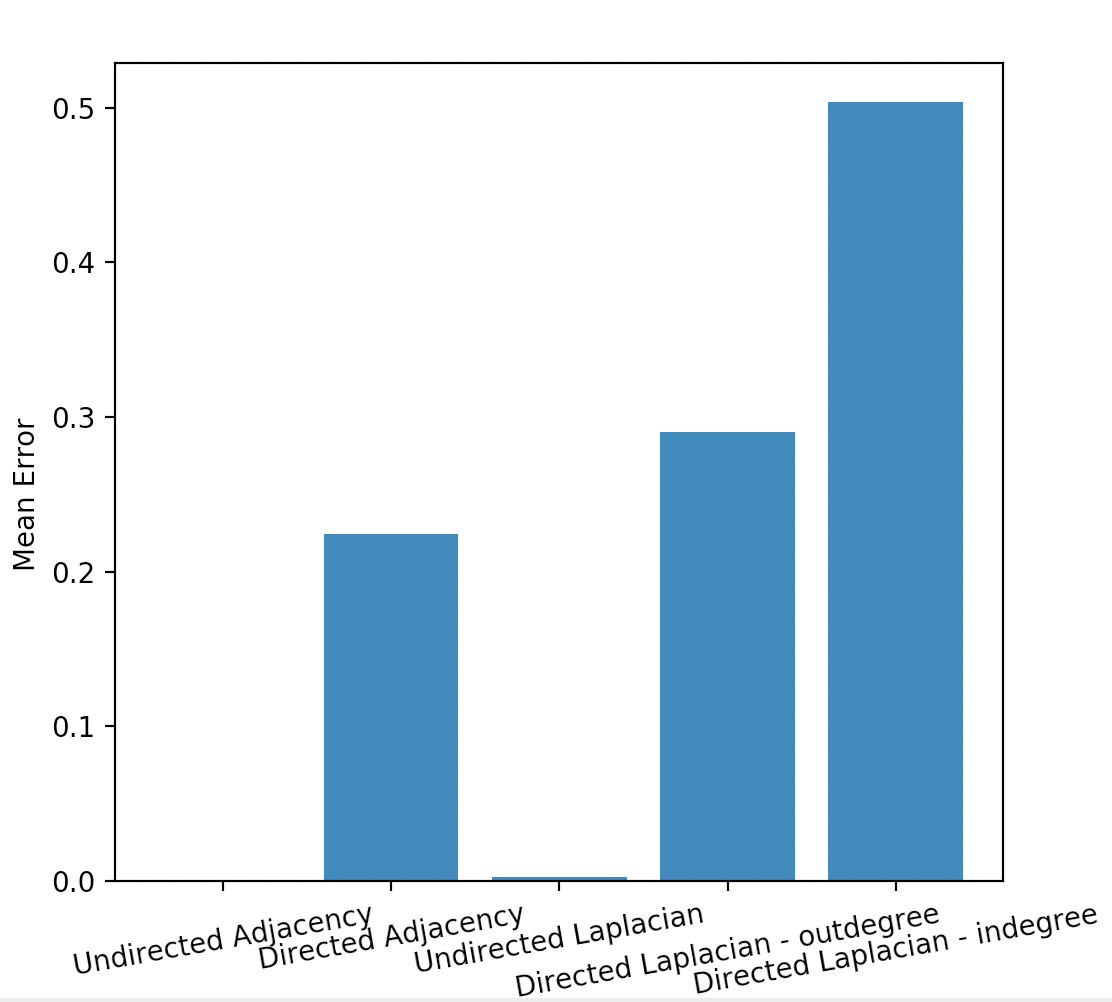}
 \caption{Matrix Type vs. Error}
    \label{fig:typee}
\end{figure}

We weren't sure how the asymmetry of directed graph matrices would affect the algorithm's performance, but tests in \ref{fig:typer} and \ref{fig:typee} showed that matrices of undirected graphs where much easier to compute quickly and accurately than their directed graph counterparts. Additionally, the algorithm seems to find calculating the spectra of adjacency matrices slightly easier than calculating the spectra of Laplacian matrices.
\subsection{Input Size}
\begin{figure}[t]
    \centering
 \includegraphics[width=0.5\textwidth]{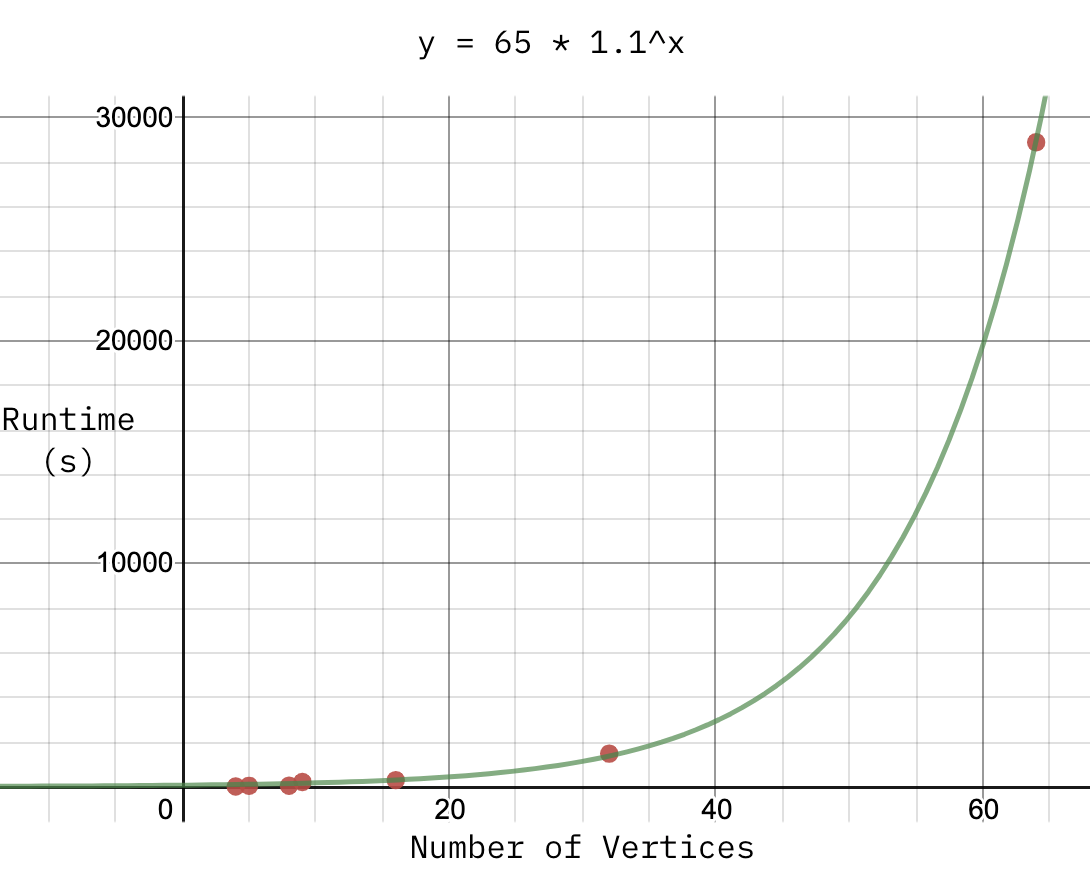}
 \caption{Exponential Curve Fit}
    \label{fig:appcurve}
\end{figure}

For our final test, we ran our algorithm on randomly generated matrices of 4, 5, 8, 9, 16, 32, and 64 vertices. The number of trials, designed for balance between variance and computational intensity, were 10, 5, 5, 2, 2, 1, and 1, respectively. 
\\
\\
\begin{center}
\begin{tabu} to 0.5\textwidth { | X[c] | X[c] | }
 \hline
 Matrix Type & Undirected Adjacency  \\
 \hline
 Density & 0.5  \\
  \hline
 Number of Ansatz Layers & 3  \\
\hline
 Lattice & Aspen-4-6Q-A  \\
\hline
\end{tabu}
\end{center}
\begin{figure}[t]
    \centering
 \includegraphics[width=0.5\textwidth]{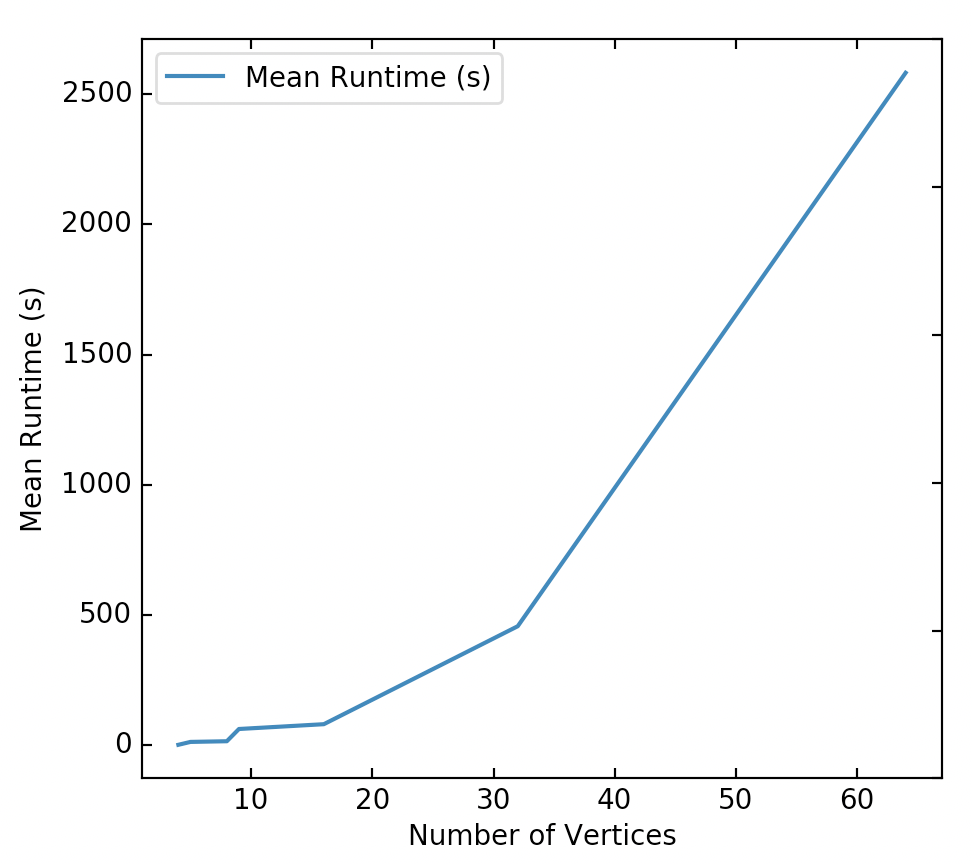}
 \caption{Input Size vs. Runtime}
    \label{fig:time}
\end{figure}
\begin{figure}[t]
    \centering
 \includegraphics[width=0.5\textwidth]{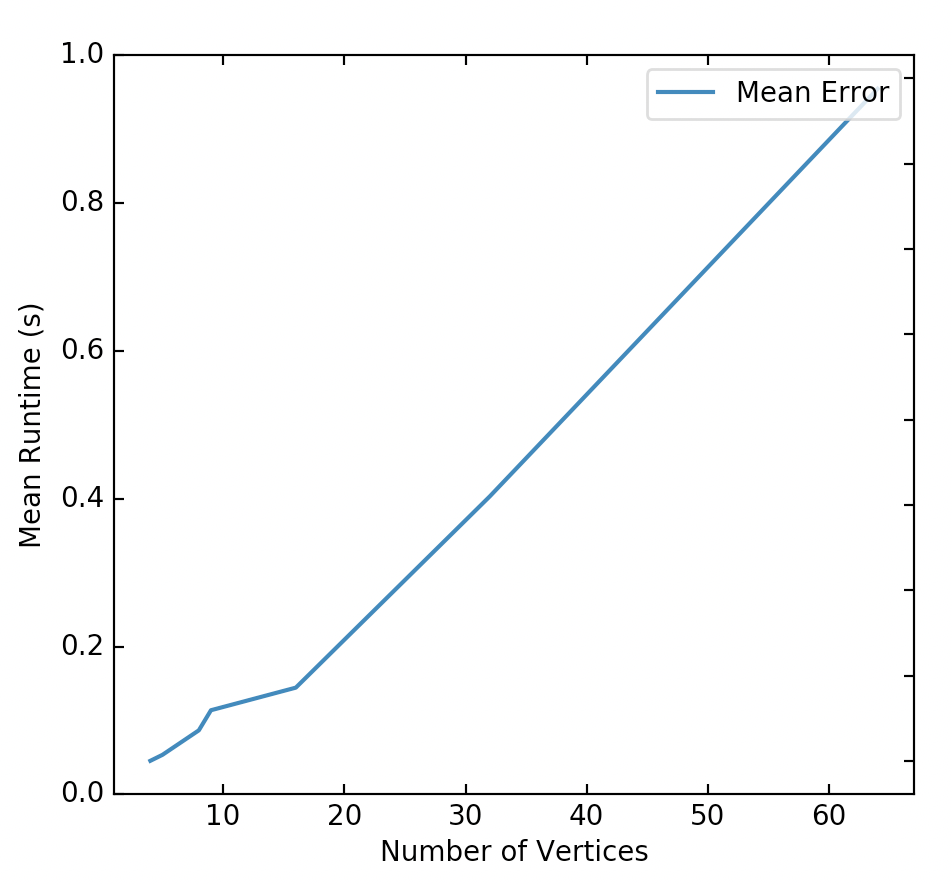}
 \caption{Input Size vs. Error}
    \label{fig:errors}
\end{figure}
With the plots in \ref{fig:time} and \ref{fig:errors}, we can see that the error rate, given the ansatz, seems to grow on the order of $O(n)$, and the runtime to grown on the order of \texttt{poly}($n$). Note also that since we pad matrices with zeros to the next power of 2, the difference in runtime between a graph on 5 vertices and a graph on 8 vertices, as well as a graph on 9 vertices and a graph on 16 vertices, is very small. Indeed, when we curve-fit the mean runtime with respect to the number of vertices (here, we plot powers of 2 for the number of vertices), we find that a quadratic fits the curve well: 
\begin{figure}[t]
    \centering
 \includegraphics[width=0.5\textwidth]{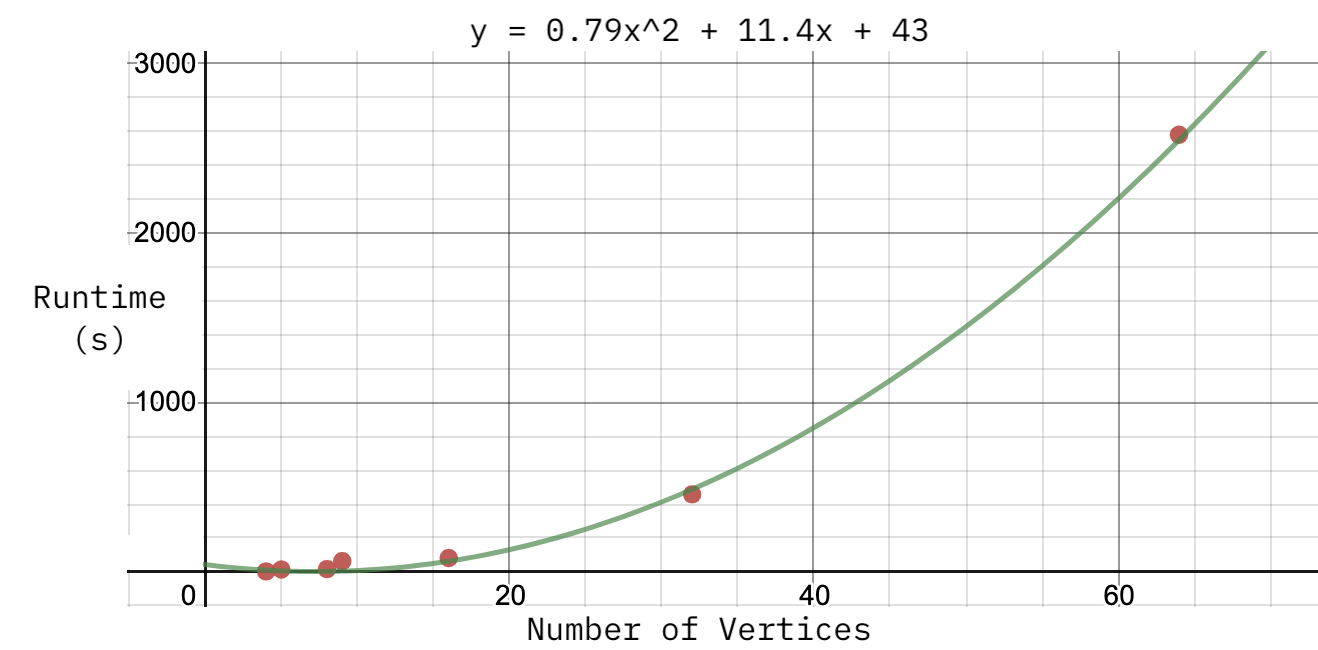}
 \caption{Polynomial Curve Fit}
    \label{fig:fit}
\end{figure}
\ref{fig:fit} seems to support the earlier theoretical claim of the runtime of each iteration being $O(n^r|h_{max}|^2M/p^2)$, with convergence of Nelder-Mead being linear. We ran this experiment on a classical machine using the quantum virtual machine, and with this experiment, we observed an exponential curve fit for the runtime plot. We find that a quadratic does not fit the curve, which seems to grow on the order of \texttt{exp}$(n)$, shown in Figure \ref{fig:appcurve}.
\newpage
.

\section{Conclusion}
In this paper, we've presented an algorithm that represents an adjacency or Laplacian matrices of graphs as a Pauli operation and applies the Variational Quantum Eigensolver (VQE) algorithm to determine the spectra of these graphs. We've discussed theoretical results regarding the runtime of this procedure and have compared these with results gathered by testing our algorithm on a quantum computer (via Rigetti's QCS). We've also observed and analyzed how our algorithm's runtime and accuracy change with respect to graph density, number of ansatz layers, graph and matrix types, and number of vertices in the graph. 

We've identified several avenues for future work. First, and perhaps most important, is the discovery of an ansatz that is particularly well suited for graph spectral matrix inputs. The current ansatz is not designed to scale, as the authors of \cite{kandala} state. Previous ansatzes have been designed with deep experience in the systems they're meant for in mind (e.g., quantum chemistry), so more work is needed here. One approach that we would like to try soon is to automate the ansatz search by either brute-force search or a statistical learning procedure. This could greatly improve the usefulness and efficacy of the variational quantum eigensolver. Second is the comparison with algorithms for determining the entire spectra of matrices. Algorithms presented in \cite{abrams} use the Quantum Fast Fourier Transform (QFFT) to provide an exponential speedup for finding eigenvalues and eigenvectors. While we've shown that our algorithm can iteratively determine all of the eigenvalues and eigenvectors of an adjacency or Laplacian matrix, it remains to see which is faster in practice. Third is theoretical work on the growth of the error rate of our algorithm with respect to the input size, as well as work on the worst- and average-case gate complexity with respect to input size. Finally, we'd like to investigate how this algorithmic approach can be applied to probabilistic methods in graph theory.

\acknowledgements

The authors acknowledge Rigetti Computing for providing quantum computing credit that was used for experiments in this work. They would also like to thank Will Zeng, Aaron Sidford, Jacob Fox, Erik Bates, and Nick Steele for their thoughts and input with the project, as well as Dan Boneh, Will Zeng, Duliger Ibeling, and Jonathan Braatz for advising them in this project.

\bibliography{report} 
\bibliographystyle{spiebib} 
\section*{Appendix}
For completeness, we simulate each of the experiments run on QCS on the quantum virtual machine (QVM). These experiments may be run on a classical computer; in this case, they were run on an macOS machine with a 2.9 GHz i9 processor and 32 GB of 2400 MHz DDR4 SDRAM without multithreading. 
\\
\subsection*{Densities}
\begin{figure}[t]
    \centering
\includegraphics[width=0.5\textwidth]{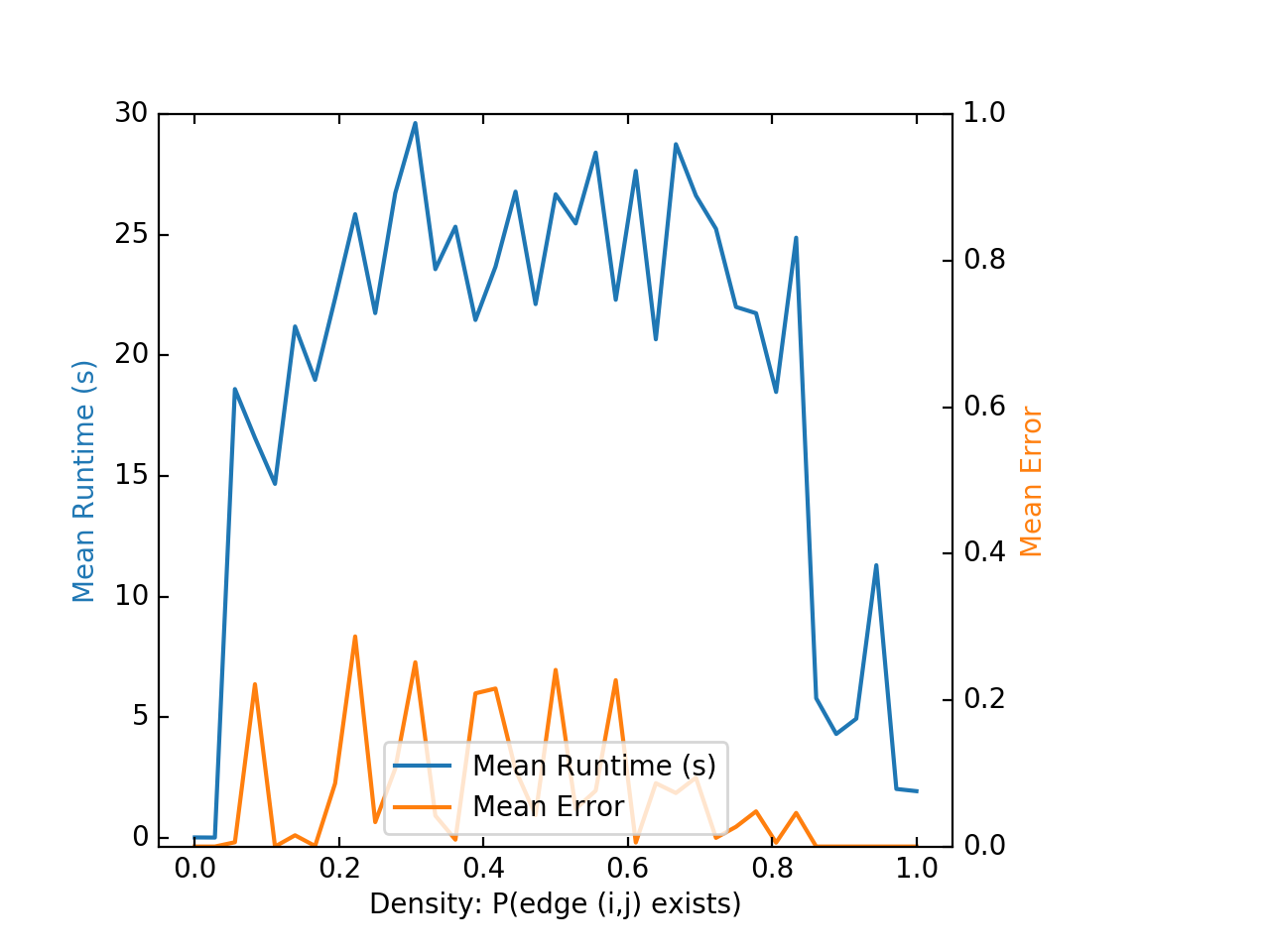}
\caption{Runtime and Error vs. Density}
    \label{fig:appdens}
\end{figure}

This experiment was carried out on 36 densities spaced uniformly between 0 and 1. The other parameters are as follows: 
\\
\\
\begin{center}
\begin{tabu} to 0.5\textwidth { | X[c] | X[c] | }
 \hline
 Matrix Type & Undirected Adjacency \\
 \hline
 Number of Vertices & 8  \\
 \hline
 Number of Trials per Test & 3  \\
 \hline
 Number of Ansatz Layers & 3  \\
\hline
\end{tabu}
\end{center}
The results are seen in Figure \ref{fig:appdens}.

\begin{figure}[t]
    \centering
 \includegraphics[width=0.5\textwidth]{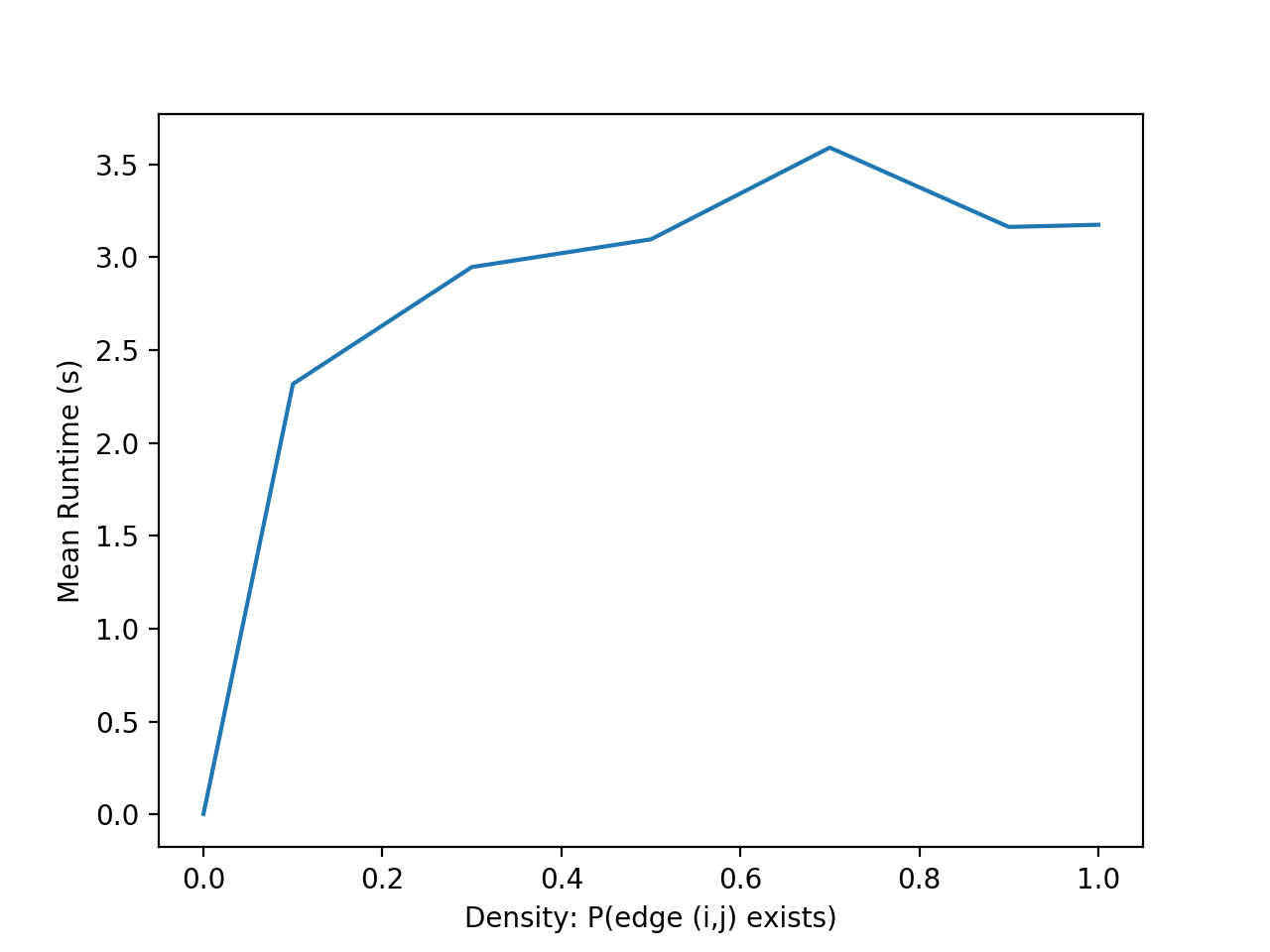}
 \caption{Runtime vs. Density}
    \label{fig:appredens}
\end{figure}

We then modeled the experiment with densities 0, 0.1, 0.3, 0.5, 0.7, 0.9, and 1 with a greater number of trials.
\\
\\
\begin{center}
\begin{tabu} to 0.5\textwidth { | X[c] | X[c] | }
 \hline
 Matrix Type & Undirected Adjacency \\
 \hline
 Number of Vertices & 4  \\
 \hline
 Number of Trials per Test & 20  \\
 \hline
 Number of Ansatz Layers & 5  \\
\hline
\end{tabu}
\end{center}
The results are seen in Figure \ref{fig:appredens}.

\subsection*{Ansatz Layers}
\begin{figure}[t]
    \centering
 \includegraphics[width=0.5\textwidth]{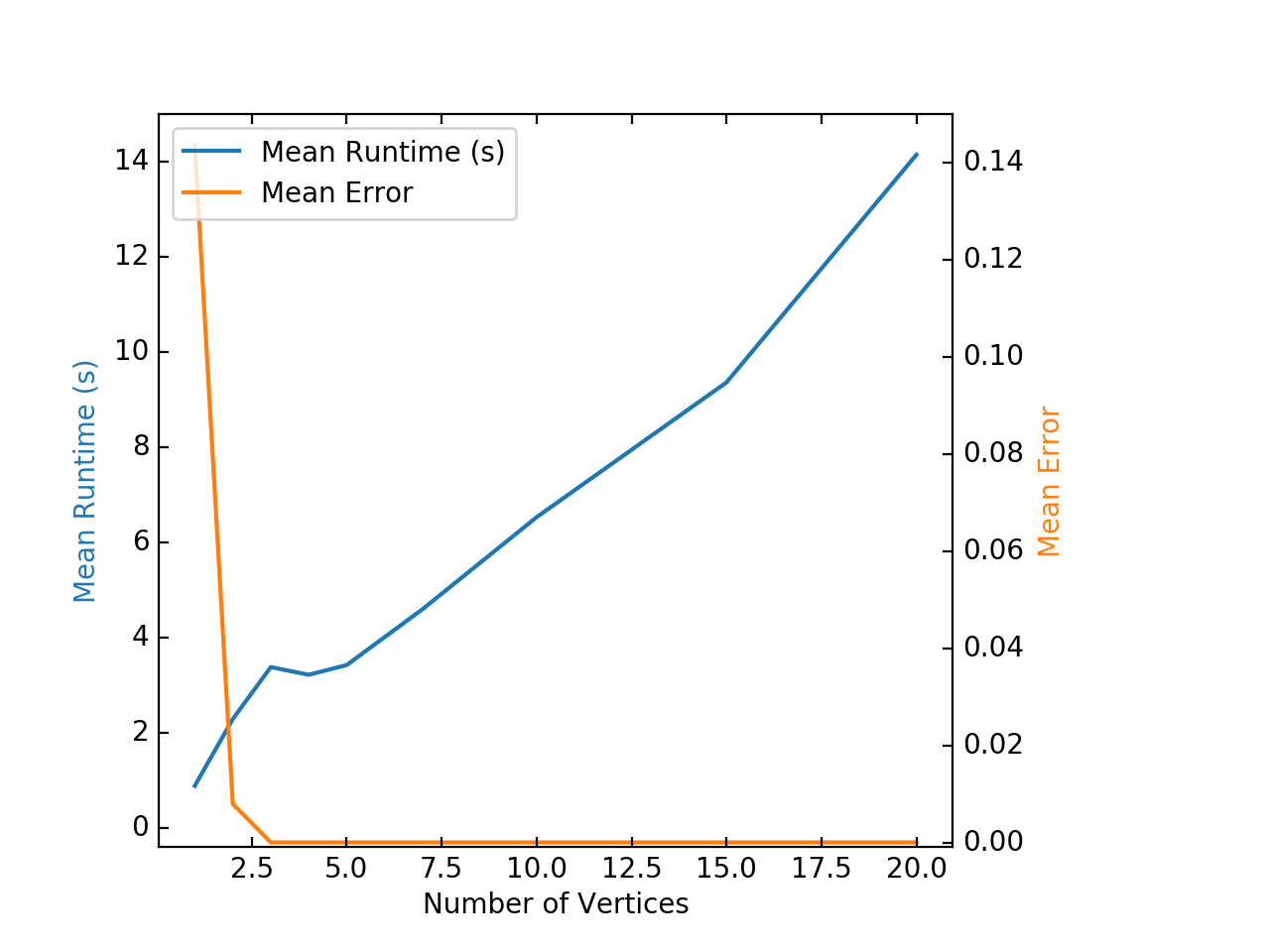} 
 \caption{Runtime vs. Ansatz Layers}
    \label{fig:appans}
\end{figure}

Next, we tested ansatzes of 1, 2, 3, 4, 5, 7, 10, 15, and 20 layers with the following parameters.
\\
\\
\begin{center}
\begin{tabu} to 0.5\textwidth { | X[c] | X[c] | }
 \hline
 Matrix Type & Undirected Adjacency \\
 \hline
 Number of Vertices & 4  \\
 \hline
 Number of Trials per Test & 20  \\
 \hline
 Density & 0.5  \\
\hline
\end{tabu}
\end{center}
The results are seen in Figure \ref{fig:appans}.

\subsection*{Graph and Matrix Types}
\begin{figure}[t]
    \centering
 \includegraphics[width=0.5\textwidth]{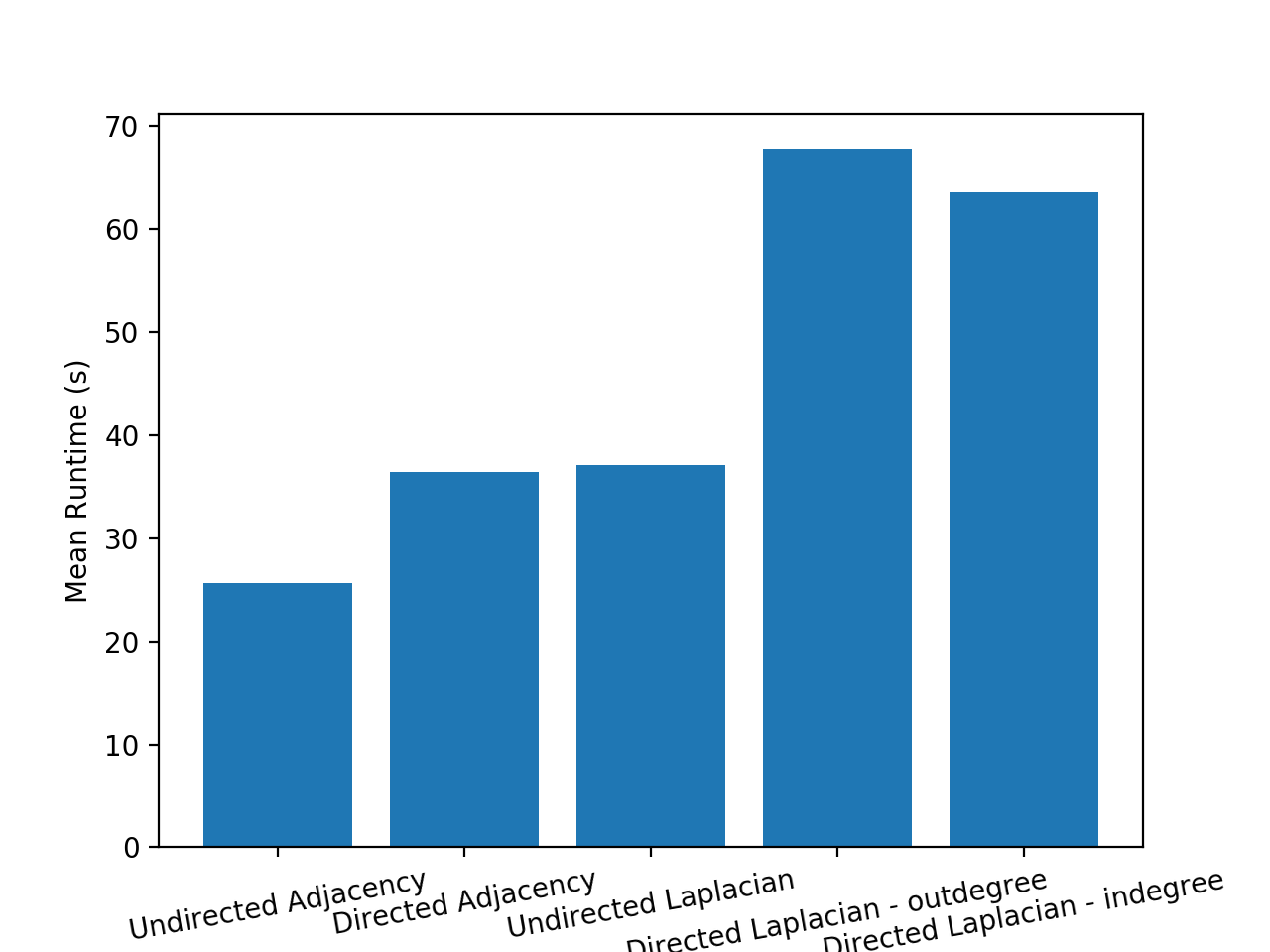}
 \caption{Matrix Type vs. Runtime}
    \label{fig:apptyper}
\end{figure}

\begin{figure}[t]
    \centering
 \includegraphics[width=0.5\textwidth]{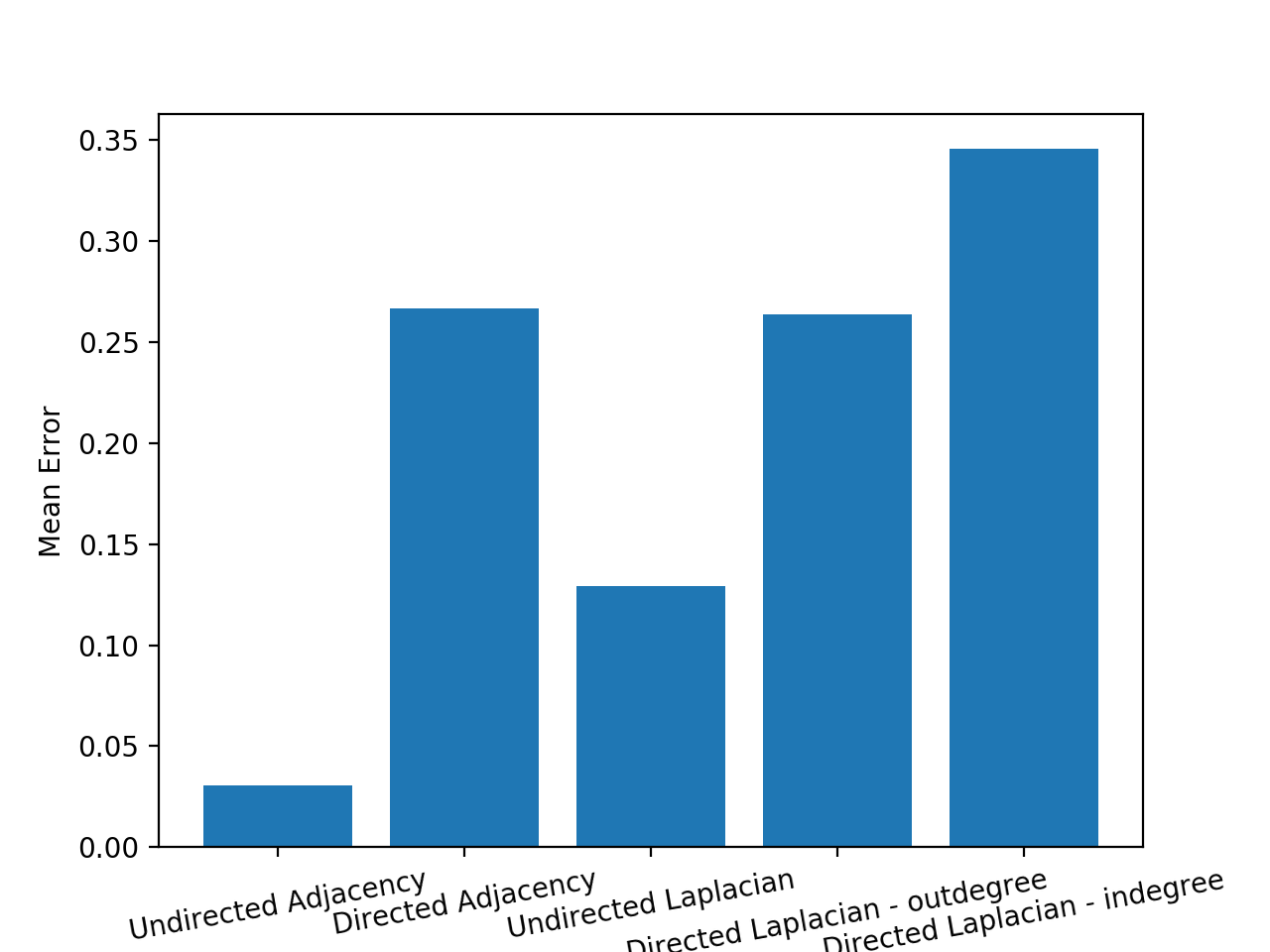}
 \caption{Matrix Type vs. Error}
    \label{fig:apptypee}
\end{figure}

Next, we experimented with the 5 aforementioned types of graphs and matrices using the following parameters. 
\\
\\
\begin{center}
\begin{tabu} to 0.5\textwidth { | X[c] | X[c] | }
 \hline
 Number of Vertices & 8  \\
 \hline
 Number of Trials per Test & 5  \\
 \hline
 Density & 0.5  \\
  \hline
 Number of Ansatz Layers & 3  \\
\hline
\end{tabu}
\end{center}
The results are seen in Figures \ref{fig:apptyper} and \ref{fig:apptypee}.

Finally, we studied the behavior of our algorithm on the QVM with respect to the number of vertices. Interestingly, the results could not be fitted well with any degree 2 polynomial, and indeed exhibited exponential behavior. We didn't anticipate the exponential factor to kick in at a number of vertices as small as this, but this experiment seems to provide evidence of this happening, which would back up our claim that our algorithm sees a superpolynomial quantum speedup over the classical method. 

We tested this on 4, 5, 8, 9, 16, 32, and 64 vertices with the following parameters: 
\\
\\
\begin{center}
\begin{tabu} to 0.5\textwidth { | X[c] | X[c] | }
 \hline
 Matrix Type & Undirected Adjacency  \\
 \hline
 Density & 0.5  \\
  \hline
 Number of Ansatz Layers & 3  \\
\hline
\end{tabu}
\end{center}
\begin{figure}[t]
    \centering
 \includegraphics[width=\textwidth]{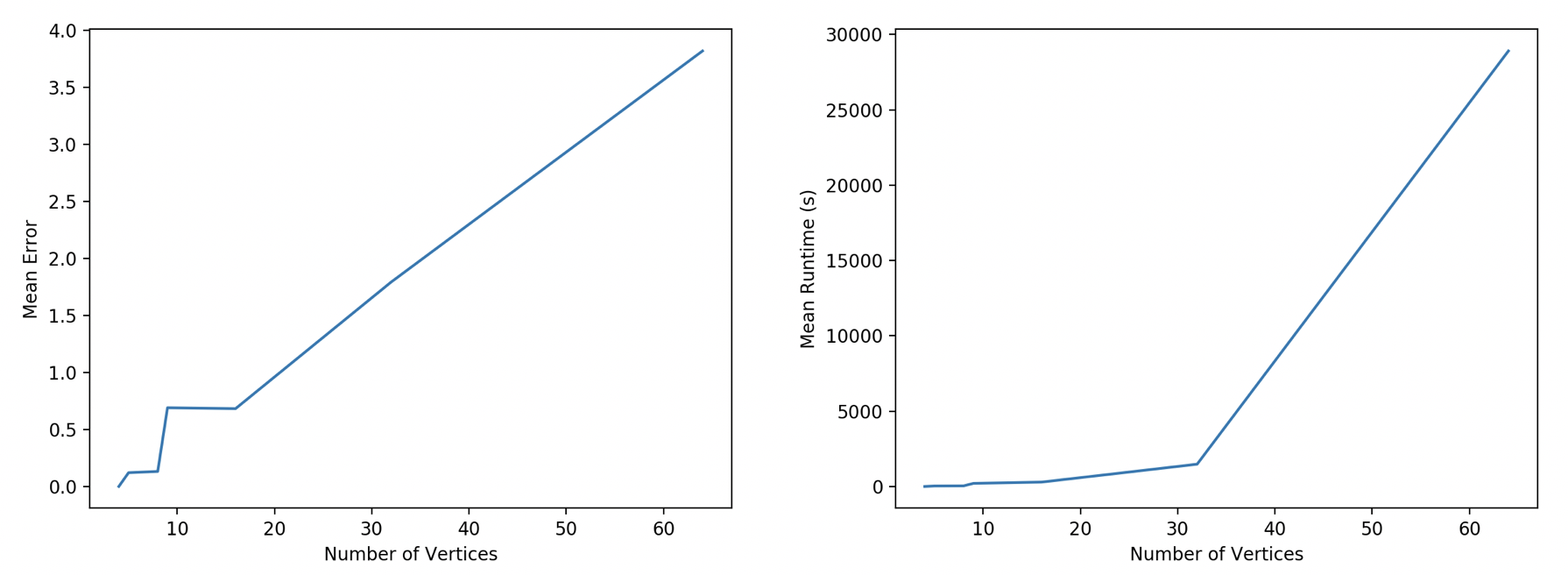}
 \caption{Input Size vs. Runtime and Input Size vs. Error}
    \label{fig:apptime}
\end{figure}

The results are seen in Figure \ref{fig:apptime}.

We see similar jumps in runtime and error from 4 to 5 and from 8 to 9 as well, since we pad an $8\times8$ matrix padded with zeros to accommodate a $5\times5$ matrix and likewise a $16\times16$ matrix padded with zeros to accommodate a $9\times9$ matrix.

\end{document}